\documentclass{llncs}

\newcommand{\keywords}[1]{\par\addvspace\baselineskip
\noindent\keywordname\enspace\ignorespaces#1}

\title{Document Clustering with K-tree}
\author{Christopher M. De Vries \and Shlomo Geva}
\institute{Faculty of Science and Technology,\\
Queensland University of Technology, Brisbane, Australia\\
\email{chris@de-vries.id.au} \hspace{10px} \email{s.geva@qut.edu.au}}

\usepackage{amsmath}
\usepackage{graphicx}
\usepackage{verbatim}
\usepackage{wrapfig}

\begin{document}

\maketitle

\begin{abstract}
This paper describes the approach taken to the XML Mining track at INEX 2008 by a group at the Queensland University of Technology. We introduce the K-tree clustering algorithm in an Information Retrieval context by adapting it for document clustering. Many large scale problems exist in document clustering. K-tree scales well with large inputs due to its low complexity. It offers promising results both in terms of efficiency and quality. Document classification was completed using Support Vector Machines.
\end{abstract}

\section{Introduction}

The XML Mining track consists of two tasks, classification and clustering. Classification labels documents in known categories. Clustering groups similar documents without any knowledge of categories. The corpus consisted of 114,366 documents and 636,187 document-to-document links. It is a subset of the XML Wikipedia corpus \cite{Denoyer2006}. Submissions were made for both tasks using several techniques.

We introduce K-tree in the Information Retrieval context. K-tree is a tree structured clustering algorithm introduced by Geva \cite{Geva2000} in the context of signal processing. It is particularly suitable for large collections due to its low complexity. Non-negative Matrix Factorization (NMF) was also used to solve the clustering task. Applying NMF to document clustering was first described by Xu et. al. at SIGIR 2003 \cite{Xu2003}. Negentropy has been used to measure clustering performance using the labels provided for documents. Entropy has been used by many researchers \cite{Surdeanu2005,Hotho2003,Steinbach2000} to measure clustering results. Negentropy differs slightly but is fundamentally measuring the same system property.

The classification task was solved using a multi-class Support Vector Machine (SVM). Similar approaches have been taken by Joachims \cite{Joachims1998} and Tong and Koller \cite{Tong2002}. We introduce a representation for links named Link Frequency Inverse Document Frequency (LF-IDF) and make several extensions to it.

Sections~\ref{sec:docRep},~\ref{sec:classTask},~\ref{sec:docClusQual},~\ref{sec:ktree},~\ref{sec:nmf} and~\ref{sec:clustTask} discuss document representation, classification, cluster quality, K-tree, NMF and clustering respectively. The paper ends with a discussion of future research and a conclusion in Sects.~\ref{sec:futureWork} and ~\ref{sec:conclusion}.

\section{Document Representation}
\label{sec:docRep}


Document content was represented with TF-IDF \cite{Salton1983} and BM25 \cite{Robertson1997}. Stop words were removed and the remaining terms were stemmed using the Porter algorithm \cite{Porter2006}. TF-IDF is determined by term distributions within each document and the entire collection. Term frequencies in TF-IDF were normalized for document length. BM25 works with the same concepts as TF-IDF except that is has two tuning parameters. The BM25 tuning parameters were set to the same values as used for TREC \cite{Robertson1997}, $K1 = 2$ and $b = 0.75$. $K1$ influences the effect of term frequency and $b$ influences document length.

Links were represented as a vector of LF-IDF weighted link frequencies. This resulted in a document-to-document link matrix. The row indicates the origin and the column indicates the destination of a link. Each row vector of the matrix represents a document as a vector of link frequencies to other documents. The motivation behind this representation is that documents with similar content will link to similar documents. For example, in the current Wikipedia both car manufacturers BMW and Jaguar link to the Automotive Industry document. Term frequencies were simply replaced with link frequencies resulting in LF-IDF. Link frequencies were normalized by the total number of links in a document.

All representations were culled to reduce the dimensionality of the data. This is necessary to fit the representations in memory when using a dense representation. K-tree will be extended to work with sparse representations in the future. A feature's rank is calculated by summation of its associated column vector. This is the sum of all weights for each feature in all documents. Only the top n features are kept in the matrix and the rest are discarded. TF-IDF was culled to the top 2000 and 8000 features. The selection of 2000 and 8000 features is arbitrary. BM25 and LF-IDF were only culled to the top 8000 features.

\section{Classification Task}
\label{sec:classTask}

The classification task was completed using an SVM and content and link information. This approach allowed evaluation of the different document representations. It allowed the most effective representation to be chosen for the clustering task. 

SVM$^\mathrm{multiclass}$ \cite{Tsochantaridis2005} was trained with TF-IDF, BM25 and LF-IDF representations of the corpus. BM25 and LF-IDF feature vectors were concatenated to train on both content and link information simultaneously. Submissions were made only using BM25, LF-IDF or both because BM25 out performed TF-IDF.

\subsection{Classification Results}

Table~\ref{table:classResults} lists the results for the classification task. They are sorted in order of decreasing recall. Recall is simply the accuracy of predicting labels for documents not in the training set. Concatenating the link and content representations did not drastically improve performance. Further work has been subsequently performed to improve classification accuracy.

\begin{table}
\begin{center}
\begin{tabular}{lclc}
\hline\noalign{\smallskip}
Name & Recall & Name & Recall \\
\noalign{\smallskip}
\hline
\noalign{\smallskip}
Expe 5 tf idf T5 10000 & 0.7876 & Expe 4 tf idf T5 100 & 0.7231 \\
Expe 3 tf idf T4 10000 & 0.7874 & Kaptein 2008NBscoresv02 & 0.6981 \\
Expe 1 tf idf TA & 0.7874 & Kaptein 2008run & 0.6979 \\
Vries text and links & 0.7849 & Romero naïve bayes & 0.6767 \\
Vries text only & 0.7798 & Expe 2.tf idf T4 100 & 0.6771 \\
Boris inex tfidf1 sim 0.38.3 & 0.7347 & Romero naïve bayes links & 0.6814 \\
Boris inex tfidf sim 037 it3 & 0.7340 & Vries links only & 0.6233 \\
Boris inex tfidf sim 034 it2 & 0.7310 & & \\
\hline
\end{tabular}
\caption{Classification Results}
\label{table:classResults}
\vspace{-20pt}
\end{center}
\end{table}

\subsection{Improving Representations}

\begin{wrapfigure}{R}{0.459\textwidth}
\vspace{-20pt}
\includegraphics[scale=0.27]{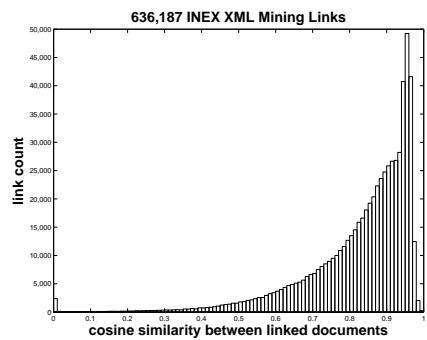}
\vspace{-20pt}
\caption{Text Similarity of Links}
\vspace{-20pt}
\label{fig:graphTextLinkHist}
\end{wrapfigure}

Several approaches have been carried out to improve classification performance. They were completed after the end of official submissions for INEX. The same train and test splits were used. All features were used for text and links, where earlier representations were culled to the top 8000 features. Links were classified without LF-IDF weighting. This was to confirm LF-IDF was improving the results. Document length normalization was removed from LF-IDF. It was noticed that many vectors in the link representation contained no features. Therefore, inbound links were added to the representation. For i, the source document and j, the destination document, a weight of one is added to the i, j position in the document-to-document link matrix. This represents an outbound link. To represent an inbound link, i is the destination document and j is the source document. Thus, if a pair of documents both link to each other they receive a weight of two in the corresponding columns in their feature vectors. Links from the entire Wikipedia were inserted into this matrix. This allows similarity to be associated on inbound and outbound links outside the XML Mining subset. This extends the 114,366$\times$114,366 document-to-document link matrix to a 114,366$\times$486,886 matrix. Classifying links in this way corresponds to the idea of hubs and authorities in HITS \cite{Kleinberg1999}. Overlap on outbound links indicates the document is a hub. Overlap on inbound links indicates the document is an authority. The text forms a 114,366$\times$206,868 document term matrix when all terms are used. The link and text representation were combined using two methods. In the first approach text and links were classified separately. The ranking output of the SVM was used to choose the most appropriate label. We call this SVM by committee. Secondly, both text and link features were converted to unit vectors and concatenated forming a 114,366$\times$693,754 matrix. Table~\ref{table:classImprovements} highlights the performance of these improvements.

\begin{table}
\begin{center}
\begin{tabular}{cllc}
\hline\noalign{\smallskip}
Dimensions & Type & Representation & Recall \\
\noalign{\smallskip}
\hline
\noalign{\smallskip}
114,366 & Links (XML Mining subset) & unweighted & 0.6874 \\
114,366 & Links (XML Mining subset) & LF-IDF & 0.6906 \\
114,366 & Links (XML Mining subset) & LF-IDF no normalization & 0.7095 \\
486,886 & Links (Whole Wikipedia) & unweighted &  0.7480 \\
486,886 & Links (Whole Wikipedia) & LF-IDF & 0.7527 \\
486,886 & Links (Whole Wikipedia) & LF-IDF no normalization & 0.7920 \\
206,868 & Text & BM25 & 0.7917 \\
693,754 & Text, Links (Whole Wikipedia) & BM25 + LF-IDF committee & 0.8287 \\
693,754 & Text, Links (Whole Wikipedia) & BM25 + LF-IDF concatenation & 0.8372 \\
\hline
\end{tabular}
\caption{Classification Improvements}
\label{table:classImprovements}
\end{center}
\end{table}

The new representation for links has drastically improved performance from a recall of 0.62 to 0.79. It is now performing as well as text based classification. However, the BM25 parameters have not been optimized. This could further increase performance of text classification. Interestingly, 97 percent of the correctly labeled documents for text and link classification agree. To further explain this phenomenon, a histogram of cosine similarity of text between linked documents was created. Figure~\ref{fig:graphTextLinkHist} shows this distribution for the links in XML Mining subset. Most linked documents have a high degree of similarity based on their text content. Therefore, it is valid to assume that linked documents are highly semantically related. By combining text and link representations we can disambiguate many more cases. This leads to an increase in performance from 0.7920 to 0.8372 recall. The best results for text, links and both combined, performed the same under 10 fold cross validation using a randomized 10\% train and 90\% test split.

LF-IDF link weighting is motivated by similar heuristics to TF-IDF term weighting. In LF-IDF the link inverse document frequency reduces the weight of common links that associate documents poorly and increases the weight of links that associate documents well. This leads to the concept of stop-links that are not useful in classification. Stop-links bare little semantic information and occur in many unrelated documents. Consider for instance a document collection of the periodic table of the elements, where each document corresponds to an element. In such a collection a link to the ``Periodic Table'' master document would provide no information on how to group the elements. Noble gases, alkali metals and every other category of elements would all link to the ``Periodic Table'' document. However, links that exist exclusively in noble gases or alkali metals would be excellent indicators of category. Year links in the Wikipedia are a good example of a stop-link as they occur with relatively high frequency and convey no information about the semantic content of pages in which they appear.


\section{Document Cluster Quality}
\label{sec:docClusQual}


The purity measure for the track is calculated by taking the most frequently occurring label in each cluster. Micro purity is the mean purity weighted by cluster size and macro is the unweighted arithmetic mean. Taking the most frequently occurring label in a cluster discards the rest of the information represented by the other labels. Due to this fact negentropy was defined. It is the opposite of information entropy \cite{Shannon1949}. If entropy is a measure of uncertainty associated with a random variable then negentropy is a measure of certainty. Thus, it is better when more labels of the same class occur together. When all labels are evenly distributed across all clusters the lowest possible negentropy is achieved.

\begin{wrapfigure}{R}{0.459\textwidth}
\vspace{-20pt}
\includegraphics[scale=0.27]{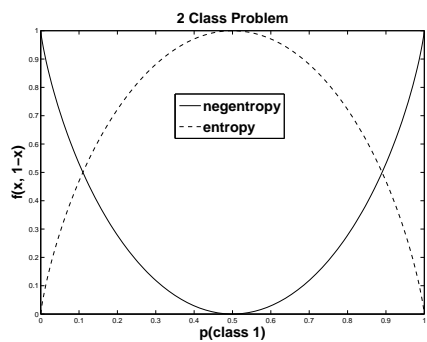}
\vspace{-20pt}
\caption{Entropy Versus Negentropy}
\vspace{-20pt}
\label{fig:graphNeg}
\end{wrapfigure}

Negentropy is defined in Equations \eqref{eq:ld}, \eqref{eq:px} and \eqref{eq:Hd}. $D$ is the set of all documents in a cluster. $X$ is the set of all possible labels.  $l(d)$ is the function that maps a document $d$ to its label $x$. $p(x)$ is the probability for label $x$. $H(D)$ is the negentropy for document cluster $D$. The negentropy for a cluster falls in the range $0 \leq H(D) \leq 1$ for any number of labels in $X$. Figure~\ref{fig:graphNeg} shows the difference between entropy and negentropy. While they are exact opposites for a two class problem, this property does not hold for more than two classes. Negentropy always falls between zero and one because it is normalized. Entropy is bounded by the number of classes. The difference between the maximum value for negentropy and entropy increase when the number of classes increase.


\begin{equation} \label{eq:ld}
l(d) = \{ (d_1,x_1), (d_2,x_2), \dots, (d_{|D|},x_{|D|}) \}
\end{equation}

\begin{equation} \label{eq:px}
p(x) = \frac{| \{ d \in D : x = l(d) \} |}{|D|}
\end{equation}

\begin{equation} \label{eq:Hd}
H(D) = 1 + \frac{1}{\log_2 |X|} \sum_{\substack{x \in X \\ p(x) \neq 0}} p(x) \log_2 p(x)
\end{equation}

\begin{figure}
 \hspace{0.033\textwidth}
 \begin{minipage}{0.4\textwidth}
  \vspace{0pt}
  \centering
  \begin{tabular}{cc}
   \hline\noalign{\smallskip}
   Cluster & Label Counts \\
   \noalign{\smallskip}
   \hline
   \noalign{\smallskip}
   1 & A=3, B=3 \\
   2 & A=3, C=3 \\
   3 & B=3, D=3 \\
   4 & C=3, D=3 \\
   \hline
  \end{tabular}
  \includegraphics[scale=0.125]{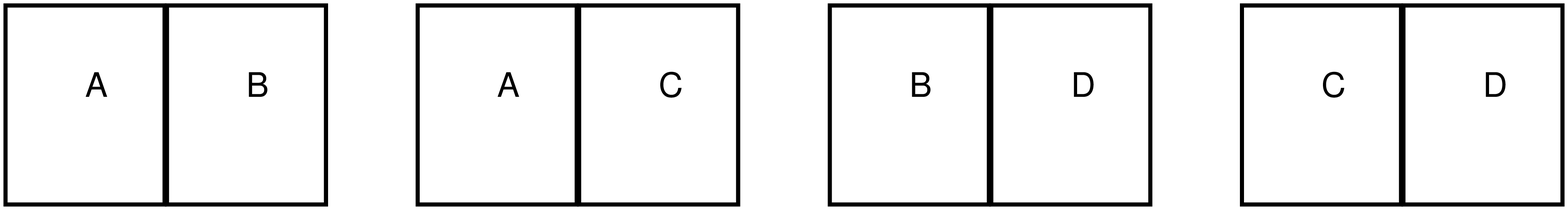}
  \caption{Solution 1}
  \label{fig:sol1}
 \end{minipage}
 \hspace{0.033\textwidth}
 \begin{minipage}{0.4\textwidth}
  \vspace{0pt}
  \centering
  \begin{tabular}{cc}
   \hline\noalign{\smallskip}
   Cluster & Label Counts \\
   \noalign{\smallskip}
   \hline
   \noalign{\smallskip}
   1 & A=3, B=1, C=1, D=1 \\
   2 & B=3, C=1, D=1, A=1 \\
   3 & C=3, D=1, A=1, B=1 \\
   4 & D=3, A=1, B=1, C=1 \\
   \hline
  \end{tabular}
  \includegraphics[scale=0.125]{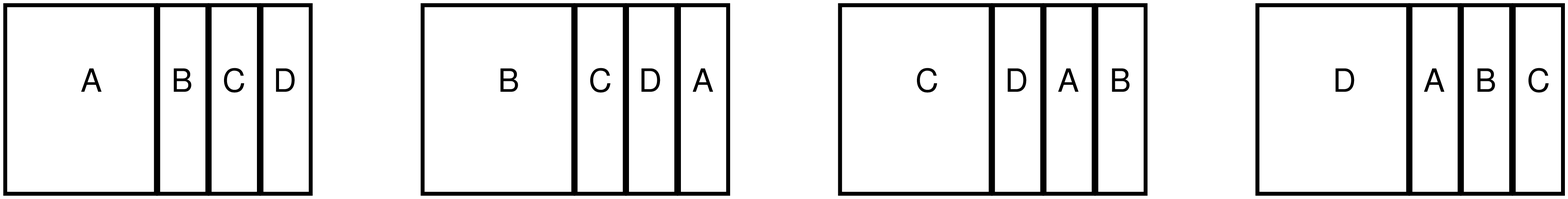}
  \caption{Solution 2}
  \label{fig:sol2}
 \end{minipage}
\end{figure}

The difference between purity and negentropy can easily be demonstrated with an artificial four class problem. There are six of each of the labels A, B, C and D. For each cluster in Solution 1 purity and negentropy is 0.5. For each cluster in Solution 2 the purity is 0.5 and the negentropy is 0.1038. Purity makes no differentiation between the two solutions. If the goal of document clustering is to group similar documents together then Solution 1 is clearly better because each label occurs in two clusters instead of four. The grouping of labels is better defined because they are less spread. Figures \ref{fig:sol1} and \ref{fig:sol2} show Solutions 1 and 2.

\section{K-tree}
\label{sec:ktree}

The K-tree algorithm is a height balanced cluster tree. It can be downloaded from \url{http://ktree.sf.net}. It is inspired by the B$^+$-tree where all data records are stored in the leaves at the lowest level in the tree and the internal nodes form a nearest neighbour search tree. The k-means algorithm is used to perform splits when nodes become full. The constraints placed on the tree are relaxed in comparison to a B$^+$-tree. This is due to the fact that vectors do not have a total order like real numbers. \\

B$^+$-tree of order $m$
\begin{enumerate}
\item All leaves are on the same level.
\item Internal nodes, except the root, contain between $\lceil \frac{m}{2} \rceil$ and $m$ children.
\item Internal nodes with $n$ children contain $n - 1$ keys, partitioning the children into a search tree.
\item The root node contains between 2 and $m$ children. If the root is also a leaf then it can contain a minimum of 0.
\end{enumerate}

K-tree of order $m$
\begin{enumerate}
\item All leaves are on the same level.
\item Internal nodes contain between one and $m$ children. The root can be empty when the tree contains no vectors.
\item Codebook vectors (cluster representatives) act as search keys.
\item Internal nodes with $n$ children contain $n$ keys, partitioning the children into a nearest neighbour search tree.
\item The level immediately above the leaves form the codebook level containing the codebook vectors.
\item Leaf nodes contain data vectors.
\end{enumerate}

The leaf nodes of a K-tree contain real valued vectors. The search path in the tree is determined by a nearest neighbour search. It follows the child node associated with nearest vector. This follows the same recursive definition of a B$^+$-tree where each tree is made up of a smaller sub tree. The current implementation of K-tree uses Euclidean distance for all measures of similarity. Future versions will have the ability to specify any distance measure.

\subsection{Building a K-tree}

\begin{figure}[!h]
\centering
\includegraphics[scale=0.137]{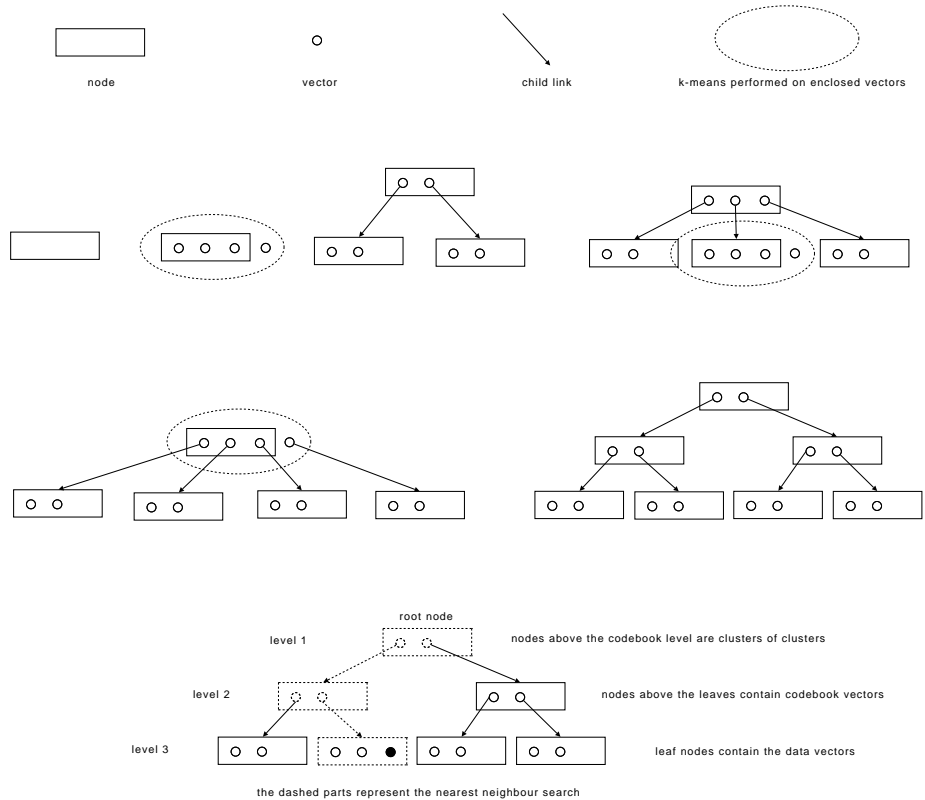}
\caption{K-tree Construction}
\label{fig:ktree2}
\end{figure}

The K-tree is constructed dynamically as data vectors arrive. Initially the tree contains a single empty root node at the leaf level. Vectors are inserted via a nearest neighbour search, terminating at the leaf level. The root of an empty tree is a leaf, so the nearest neighbour search terminates immediately, placing the vector in the root. When $m + 1$ vectors arrive the root node can not contain any more keys. It is split using k-means where $k = 2$ using all $m + 1$ vectors. The two centroids that result from k-means become the keys in a new parent. New root and child nodes are constructed and each centroid is associated with a child. The vectors associated with each centroid from k-means are placed into the associated child. This process has created a new root for the tree. It is now two levels deep. The root has two keys and two children, making a total of three nodes in the tree. Now that the tree is two levels deep, the nearest neighbour search finds the closest centroid in the root and inserts it in the associated child. When a new vector is inserted the centroids are updated along the nearest neighbour search path. They are weighted by the number of data vectors contained beneath them. This process continues splitting leaves until the root node becomes full. K-means is run on the root node containing centroids. The keys in the new root node become centroids of centroids. As the tree grows internal and leaf nodes are split in the same manner. The process can potentially propagate to a full root node and cause construction of a new root. Figure~\ref{fig:ktree2} shows this construction process for a K-tree of order three ($m = 3)$.

\begin{wrapfigure}{R}{0.459\textwidth}
\vspace{-20pt}
\includegraphics[scale=0.27]{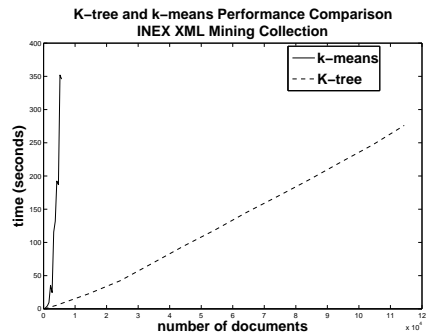}
\vspace{-20pt}
\caption{K-tree Performance}
\vspace{-20pt}
\label{fig:performance}
\end{wrapfigure}

The time complexity of building a K-tree for $n$ vectors is O($n$ log $n$). An insertion of a single vector has the time complexity of O(log $n$). These properties are inherent to the tree based algorithm. This allows the K-tree to scale efficiently with the number of input vectors. When a node is split, k-means is always restricted to $m + 1$ vectors and two centroids ($k = 2$). Figure~\ref{fig:performance} compares k-means performance with K-tree where $k$ for k-means is determined by the number of codebook vectors. This means that both algorithms produce the same number of document clusters and this is necessary for a meaningful comparison.  The order, $m$, for K-tree was 50. Each algorithm was run on the 8000 dimension BM25 vectors from the XML mining track.

\subsection{K-tree Submissions}

K-tree was used to create clusters using the Wikipedia corpus. Documents were represented as 8000 dimension BM25 weighted vectors. Thus, clusters were formed using text only. This representation was used because it was most effective text representation in the classification task. The K-tree was constructed using the entire collection. Cluster membership was determined by comparing each document to all centroids using cosine similarity. The track required a submission with 15 clusters but K-tree does not produce a fixed number of clusters. Therefore, the codebook vectors were clustered using k-means++ where $k = 15$. The codebook vectors are the cluster centroids that exist above the leaf level. This reduces the number of vectors used for k-means++, making it quick and inexpensive. As k-means++ uses a randomised seeding process, it was run 20 times to find the solution with the lowest distortion. The k-means++ algorithm \cite{Arthur2007} improves k-means by using the D$^2$ weighting for seeding. Two other submission were made representing different levels of a K-tree. A tree of order 100 had 42 clusters in the first level and a tree of order 20 had 147 clusters in the second level. This made for a total of three submissions for K-tree.

\begin{wrapfigure}{R}{0.459\textwidth}
\vspace{-20pt}
\includegraphics[scale=0.27]{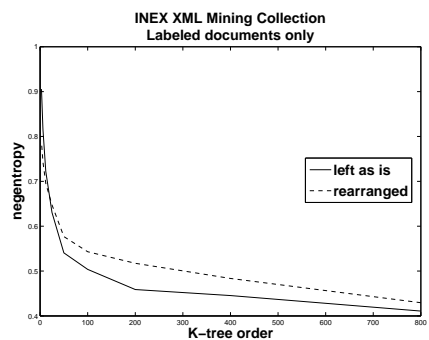}
\vspace{-20pt}
\caption{K-tree Negentropy}
\vspace{-40pt}
\label{fig:ktreeneg}
\end{wrapfigure}

Negentropy was used to determine the optimal tree order. K-tree was built using the documents in the 10\% training set from the classification task. A tree was constructed with an order of 800 and it was halved each time. Negentropy was measured in the clusters represented by the leaf nodes. As the order decreases the size of the nodes shrinks and the purity increases. If all clusters became pure at a certain size then decreasing the tree order further would not improve negentropy. However, this was not the case and negentropy continued to increase as the tree order decreased. This is expected because there will usually be some imperfection in the clustering with respect to the labels. Therefore, the sharp increase in negentropy in a K-tree below an order of 100 suggests that the natural cluster size has been observed. This can be seen in Fig.~\ref{fig:ktreeneg}. The ``left as is'' line represents the K-tree as it is built initially. The ``rearranged'' line represents the K-tree when all the leaf nodes have been reinserted to their nearest neighbours without modifying the internal nodes.

Negentropy was calculated using the 10\% training set labels provided on clusters for the whole collection. This was used to determine which order of 10, 20 or 35 fed into k-means++ with $k = 15$ was best. A tree of order 20 provided the best negentropy.

\section{Non-negative Matrix Factorization}
\label{sec:nmf}

NMF factorizes a matrix into two matrices where all the elements are $\geq 0$. If $V$ is a $n \times m$ matrix and $r$ is a positive integer where $r < min(n,m)$, NMF finds two non-negative matrices $W_{n \times r}$ and $H_{r \times m}$ such that $V \approx WH$. When applying this process to document clustering $V$ is a term document matrix. Each column in $H$ represents a document and the largest value represents its cluster. Each row in $H$ is a cluster and each column is a document.

The projected gradient method was used to solve the NMF problem \cite{Lin2007}. $V$ was a $8000 \times 114366$ term document matrix of BM25 weighted terms. The algorithm ran for a maximum of 70 iterations. It produced the $W$ and $H$ matrices. Clusters membership was determined by the maximum value in the columns of $H$. NMF was run with $r$ at 15, 42 and 147 to match the submissions made with K-tree.

\section{Clustering Task}
\label{sec:clustTask}

\begin{table}
\begin{center}
\begin{tabular}{lccclccc}
\hline\noalign{\smallskip}
Name & Size & Micro & Macro & Name & Size & Micro & Macro\\
\noalign{\smallskip}
\hline
\noalign{\smallskip}
K-tree & 15 & 0.4949 & 0.5890 & QUT LSK 1 & 15 & 0.4518 & 0.5594 \\
QUT LSK 3 & 15 & 0.4928 & 0.5307 & QUT LSK 4 & 15 & 0.4476 & 0.4948 \\
QUT Entire collection 15 & 15 & 0.4880 & 0.5051 & QUT LSK 2 & 15 & 0.4442 & 0.5201 \\
NMF & 15 & 0.4732 & 0.5371 & Hagenbuchner & 15 & 0.3774 & 0.2586 \\
\hline
\end{tabular}
\caption{Clustering Results Sorted by Micro Purity}
\end{center}
\end{table}

\begin{table}
\begin{center}
\begin{tabular}{lccccccccc}
\hline\noalign{\smallskip}
Method & Clusters & Micro & Macro & Clusters & Micro & Macro & Clusters & Micro & Macro \\
\noalign{\smallskip}
\hline
\noalign{\smallskip}
left as is & 17 & 0.4018 & 0.5945 & 397 & 0.5683 & 0.6996 & 7384 & 0.6913 & 0.7202 \\
rearranged & 17 & 0.4306 & 0.6216 & 371 & 0.6056 & 0.7281 & 5917 & 0.7174 & 0.7792 \\
cosine & 17 & 0.4626 & 0.6059 & 397 & 0.6584 & 0.7240 & 7384 & 0.7437 & 0.7286 \\
\hline
\end{tabular}
\caption{Comparison of Different K-tree Methods}
\label{tab:ktreeComparison}
\end{center}
\end{table}

\begin{figure}
\begin{center}
 \centerline{
    \mbox{\includegraphics[scale=0.29]{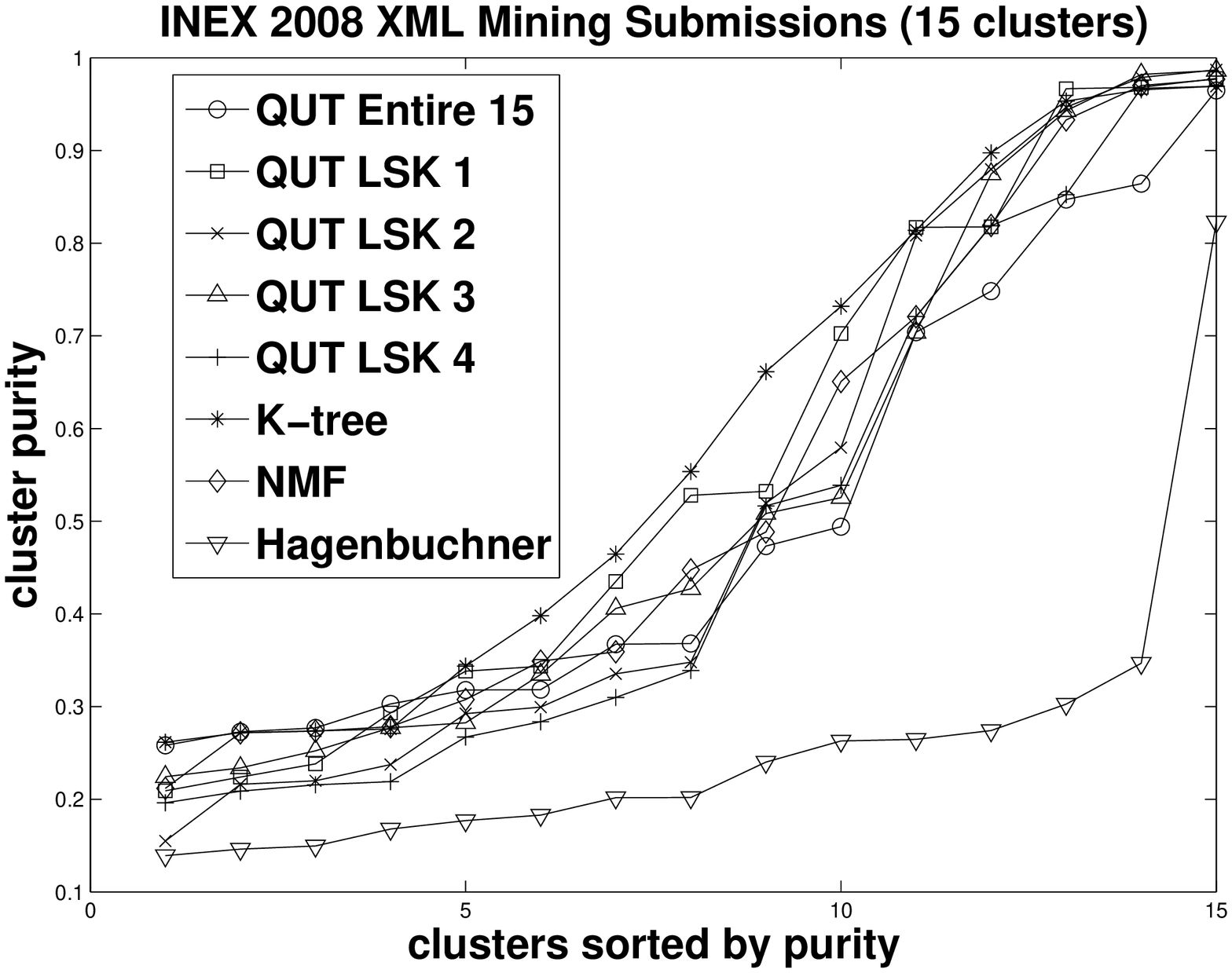}}
    \mbox{\includegraphics[scale=0.29]{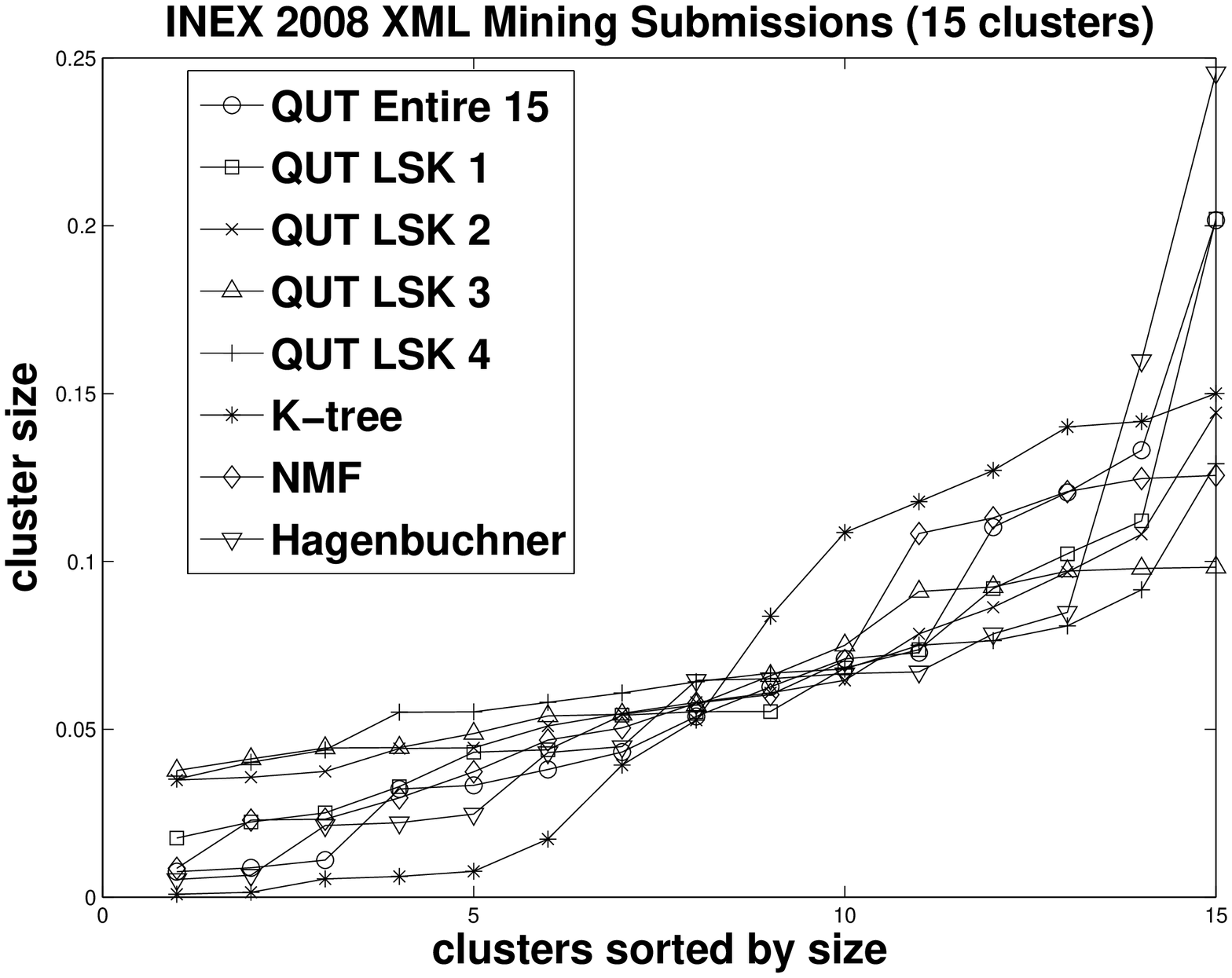}}
 }
\caption{All Submissions with 15 Clusters}
\label{fig:15all}
\end{center}
\begin{center}
\includegraphics[scale=0.29]{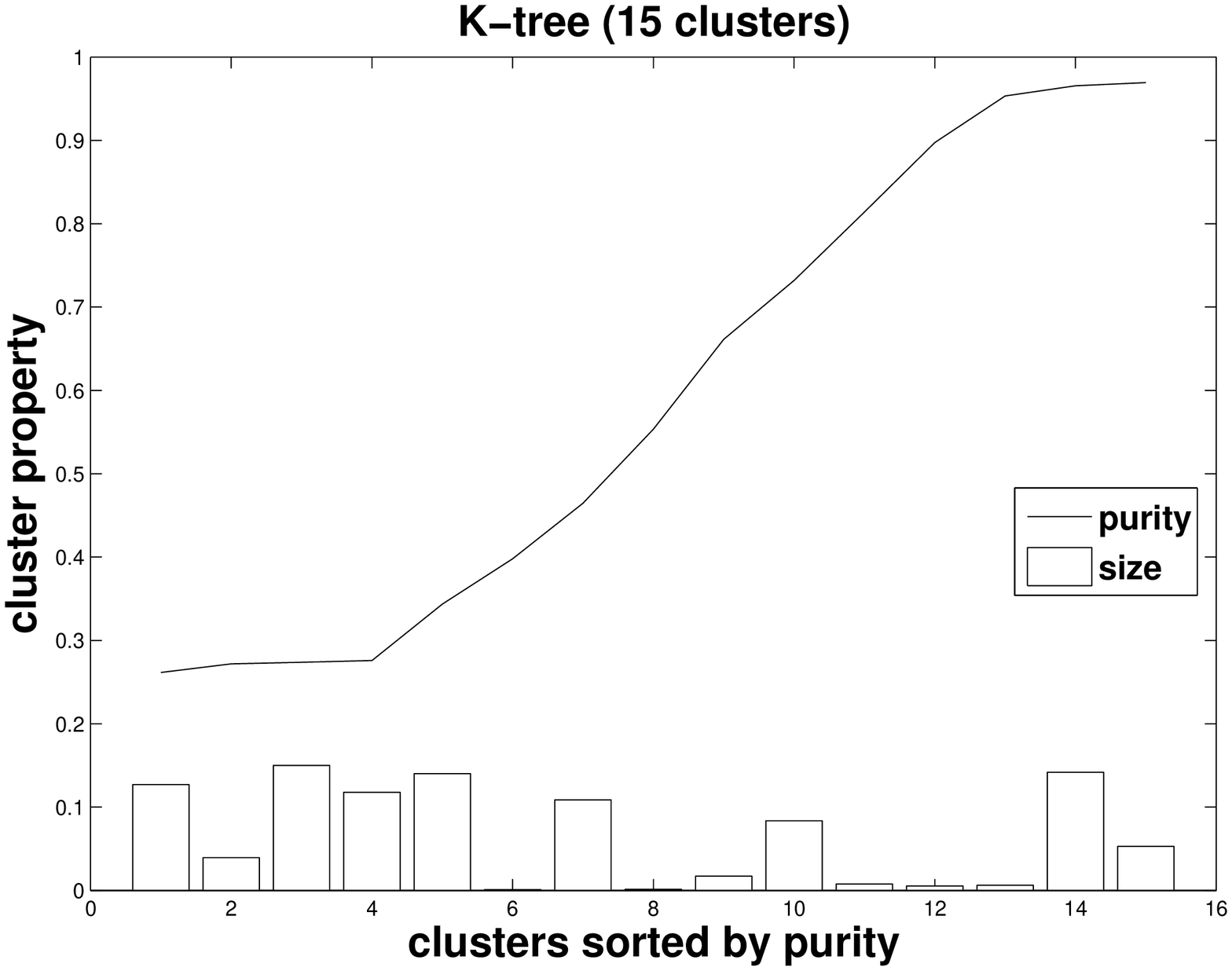}
\caption{K-tree Breakdown}
\label{fig:15breakdown}
\end{center}
\end{figure}

Every team submitted at least one solution with 15 clusters. This allows for a direct comparison between different approaches. It only makes sense to compare results where the number of clusters are the same. The K-tree performed well according to the macro and micro purity measures in comparison to the rest of the field. The difference in macro and micro purity for the K-tree submissions can be explained by the uneven distribution of cluster sizes. Figure~\ref{fig:15breakdown} shows that many of the higher purity clusters are small. Macro purity is simply the average purity for all clusters. It does not take cluster size into account. Micro purity does take size into account by weighting purity in the average by the cluster size. Three types of clusters appear when splitting the x-axis in Fig.~\ref{fig:15breakdown} in thirds. There are very high purity clusters that are easy to find. In the middle there are some smaller clusters that have varying purity. The larger, lower purity clusters in the last third are hard to distinguish. Figure~\ref{fig:15all} shows clusters sorted by purity and size. K-tree consistently found higher purity clusters than other submissions. Even with many small high purity clusters, K-tree achieved a high micro purity score. The distribution of cluster size in K-tree was less uniform than other submissions. This can be seen in Figure~\ref{fig:15all}. It found many large clusters and many small clusters, with very few in between.

The K-tree submissions were determined by cosine similarity with the centroids produced by K-tree. The tree has an internal ordering of clusters as well. A comparison between the internal arrangement and cosine similarity is listed in Table~\ref{tab:ktreeComparison}. This data is based on a K-tree of order 40. Levels 1, 2 and 3 produced 17, 397 and 7384 clusters respectively. Level 3 is the above leaf or codebook vector level. The ``left as is'' method uses the K-tree as it is initially built. The rearranged method uses the K-tree when all vectors are reinserted into the tree to their nearest neighbour. The cosine method determines cluster membership by cosine similarity with the centroids produced. Nodes can become empty when reinserting vectors. This explains why levels 2 and 3 in the rearranged K-tree contain less clusters. Using cosine similarity with the centroids improved purity in almost all cases.

\section{Future Work}
\label{sec:futureWork}

The work in this area falls into two categories, XML mining and K-tree. Further work in the XML mining area involves better representation of structure. For example, link information can be included into clustering via a modified similarity measure for documents. Future work with the K-tree algorithm will involve different strategies to improve quality of clustering results. This will require extra maintenance of the K-tree as it is constructed. For example, reinsertion of all data records can happen every time a new root is constructed.

\section{Conclusion}
\label{sec:conclusion}

In this paper an approach to the XML mining track was presented, discussed and analyzed. A new representation for links was introduced, extended and analyzed. It was combined with text to further improve classification performance. The K-tree algorithm was applied to document clustering for the first time. The results show that it is well suited for the task. It produces good quality clustering solutions and provides excellent performance.

\bibliographystyle{splncs}
\bibliography{bib}

\end{document}